# QUANTUM ZENO, HAMLET AND SIMILAR EFFECTS

Vladan Panković

Department of Physics, Faculty of Sciences, 21000 Novi Sad, Trg Dositeja Obradovića 4
Serbia, vladan.pankovic@df.uns.ac.rs

## Abstract

Simply speaking quantum Zeno effect for an unstable quantum system represents total decay probability decrease by frequent decay detections. Analogously simply speaking quantum anti-Zeno effect for an unstable quantum system represents total no-decay probability decrease by frequent decay detections. Also, some authors considered optical, i.e. polarization detection analogy of quantum Zeno and anti-Zeno effects. In this work we consider basic standard quantum mechanical concepts in quantum Zeno and anti-Zeno effect. We use simplest (Malus law) form of mentioned ideas on the photon polarization detection analogy of quantum Zeno and anti-Zeno effects and we suggest some new situations (non-equivalent time durations of the quantum dynamical evolutions between different frequent detections). Correspondingly we define new quantum effects, quantum Hamlet effect (when, by frequent detections, final total polarization angle and state converge but total probability for appearance of this state diverges) and quantum anti-Hamlet effect (when, by frequent detections, final total polarization angle and state diverge but total probability for appearance of this state converges). Also, we consider a rough classical mechanical analogy of Zeno effect. Finally we consider frequent interference of single photon at beam splitters and obtain for different values of reflection and transmission coefficients analogies of quantum Zeno and anti-Zeno effect as well as analogy of usual exponentially decreasing decay law.

## 1. Introduction

. Simply speaking quantum Zeno effect for an unstable quantum system represents total decay probability decrease by frequent decay detections [1]. Analogously simply speaking quantum anti-Zeno effect for an unstable quantum system represents total no-decay probability decrease by frequent decay detections [2]. Also, some authors [3]. [4] considered optical, i.e. polarization detection analogy of quantum Zeno and anti-Zeno effects. In this work we shall consider basic standard quantum mechanical concepts (characteristics of the quantum dynamical evolution and measurement or detection) as well as additional suppositions (equivalent time durations of the quantum dynamical evolutions between all frequent detections) in quantum Zeno and anti-Zeno effect. We shall use simplest (Malus law) form of mentioned ideas on the optical, i.e. polarization detection analogy of quantum Zeno and anti-Zeno effects and we shall suggest some new situations (non-equivalent time durations of the quantum dynamical evolutions between different frequent detections). Correspondingly we shall define new quantum effects, quantum Hamlet effect [5], when, by frequent detections, final total polarization angle and state converge but total probability for appearance of this state diverges. Also we shall define quantum anti-Hamlet effect, when, by frequent detections, final total polarization angle and state diverge but total probability for appearance of this state converges. (By simple analogy we define quantum anti-Hamlet effect for unstable quantum system too.) .) Also, we shall consider a rough classical mechanical analogy of Zeno effect. Finally we shall consider frequent interference of single photon at beam splitters and obtain for different values of reflection and transmission coefficients analogies of quantum Zeno and anti-Zeno effect as well as analogy of usual exponentially decreasing decay law.

## 2. Quantum Zeno effect

As it is well-known within standard quantum mechanical formalism [6]- [8] Schrödinger equation

(2.1) $\mathbf{H}| \Psi(t) > \equiv i\hbar d/dt \,| \Psi(t) >$

for initial quantum state

(2.2) $| \Psi(0) > = | \Psi_0>$

represents exact differential form of the exact dynamical equation of a quantum system. Here **H** represents Hamiltonian observable of given system, t – time moment, ħ- reduced Planck constant, and, $| \Psi(t) >$ – exact quantum mechanical dynamical state of given system in time moment t.

General solution of (2.1), (2.2), or integral form of exact quantum dynamics of quantum system, yields strictly deterministically quantum mechanical dynamical state in any time moment. It, formally, can be presented by

(2.3) $| \Psi(t) > \equiv \mathbf{U}(t) \,| \Psi_0> = \exp[\mathbf{H}t/(i\hbar)] \,| \Psi_0>$ .

where

(2.4) $\mathbf{U}(t) = \exp[\mathbf{H}t/(i\hbar)]$

represents formally unitary quantum mechanical dynamical evolution operator.

It is very important to be pointed out that in general case in limit when t tends to zero unitary evolution operator $\mathbf{U}(t)$ tends to unit operator 1 .

For sufficiently small t expression (2.4) can be Taylor expanded in second order approximation

(2.5) $\mathbf{U}(t) = 1 - i\,\mathbf{H}t/\hbar - \mathbf{H}^2 t^2/(2\hbar^2)$

which implies in the same approximation

(2.6) $|\,\Psi(t)> \equiv (1 - i\,\mathbf{H}t/\hbar - \mathbf{H}^2 t^2/(2\hbar^2))\,|\,\Psi_0>$ .

Suppose that in time moment t at quantum system described by (2.3) measurement or detection of such observable with $|\,\Psi_0>$ as eigen state is realized during some finite time interval $\tau$ . Then, according to standard quantum mechanical formalism, this eigen state can be detected after detection with probability

(2.7) $P(t) = |<\Psi_0\,|\,\Psi(t)>|^2$

(in further text we shall consider only such situations when expression (2.7) is not exactly 1 , i.e. when initial quantum state is not eigen state of **H**) which in mentioned second order Taylor expansion approximation turns out in

(2.8) $P(t) = 1 - \Delta\mathbf{H}^2 t^2/\hbar^2$

where $\Delta\mathbf{H}$ represents standard deviation of **H** in $|\,\Psi_0>$.

(2.9) $\Delta\mathbf{H} = |<\Psi_0\,|\,\mathbf{H}^2\,|\,\Psi_0> - |<\Psi_0\,|\,\mathbf{H}\,|\,\Psi_0>^2)^{1/2}$ .

It can be observed that (2.5) and (2.8) are satisfied for

(2.10) $\Delta\mathbf{H}\,t \ll \hbar$

which in common with Heisenberg energy-time uncertainty relation

(2.11) $\Delta\mathbf{H}\,\tau \geq \hbar/2$

implies

(2.12) $t \ll \tau$ .

Suppose now that at the same quantum system n detections of the same observable are realized, first one in time moment $t_1$ after initial time moment, second one in time moment $t_2$ after first detection etc. n-th detection in time moment $t_n$ after (n-1) – th detection, where n represents some natural number. Then probability that quantum system after all n detections, i.e. in time moment

(2.13) $t(n) = t_1 + t_2 + \dots + t_n$

or more accurately in time moment

(2.14) $T(n) = t(n) + n\,\tau$

(which includes duration of all n detections), stands in the initial quantum state equals

(2.15) $P(t_1)\,P(t_2)\dots P(t_n)$ .

As it is not hard to see, without additional suppositions, expression (2.15) in limit when n tends to infinity must tend to zero. It stands satisfied even in case when all time moments $t_1, t_2, \dots, t_n$ are sufficiently small that mentioned second order approximation of Taylor expansion can be realized and when (2.15) turns out in

(2.16) $(1 - \Delta\mathbf{H}^2\, t^2{}_1/\hbar^2)\,(1 - \Delta\mathbf{H}^2\, t^2{}_2/\hbar^2)\dots(1 - \Delta\mathbf{H}^2\, t^2{}_n/\hbar^2)$ .

Generally speaking frequent detections always imply change of the initial quantum state.

However, it can be additionally supposed that sufficiently small all time moments $t_1, t_2, \dots, t_n$ represent the following functions of n

(2.17) $t_1 = t_2 = \dots = t_n = t/n$

so that

(2.18) $t(n) = t$

and

(2.19) $T(n) = t + n\,\tau$ .

(Expression (2.19) clearly points out that even in case when t is sufficiently small and n independent T(n) linearly increases with n.) Then (2.16) turns out in

(2.20) $(1 - \Delta\mathbf{H}^2 (t/n)^2 / \hbar^2)^n = (1 - \Delta\mathbf{H}^2 t^2/(n\hbar)^2)^n$ .

Expression (2.20) tends to 1 in limit when n tends to infinity, i.e.

(2.21) $(1 - \Delta\mathbf{H}^2 t^2/(n\hbar)^2)^n \rightarrow 1$ for $n \rightarrow \infty$

which represents quantum Zeno effect. Strictly speaking here frequent detections imply that any evolution time (2.17) tends to zero and corresponding evolution operator - to unit operator.

It is obvious that quantum Zeno effect represents a general consequence of the standard quantum mechanical formalism, i.e. general consequence of the quantum mechanical dynamical evolution (in second order approximation) and frequently repeated quantum measurement or detection procedure. However it is usual (seemingly most paradoxical) that quantum Zeno effect be defined for dynamically non-stable, decay able quantum systems that initially are in non-decayed quantum state. Then, by frequent detections, quantum system stands always in non-decayed state, what is of course general consequence of the standard quantum mechanical formalism. We can remember that within standard quantum mechanical formalism there is well-known other paradox, paradox of the excited quantum states. Namely within standard quantum mechanical formalism all excited quantum states of the quantum mechanical Hamiltonian observable are dynamically non-stable and mentioned excited states decay during time (by spontaneous emission of the photons) in dynamically stable and non-decay able ground quantum state. This seemingly paradoxical situation is simply solved by extension of the standard quantum mechanical formalism (including time dependent perturbation theory) in quantum electrodynamics and quantum field theory. Quantum Zeno effect exists even within quantum electrodynamics or quantum field theory of nuclear interactions but, for reason of (2.12) and usual detection procedures characteristics, its experimental verification is technically extremely hard.

Mathematical probability structure of quantum Zeno effect is very interesting. Here, on the one hand, small "decay" parameter (t), can be divided in n equal smaller parameters whose sum is n independent. On the other hand "decay" probability for any smaller "decay" parameter decrease proportionally to square of n. It necessarily implies that conditional probability for total "no-decay" tends toward 1 in limit when n tends toward infinity.

Mentioned mathematical-probabilistic conditions can be more or less satisfactorily applied in some other even classical mechanical physical systems.

There are different classical mechanical analogies of the quantum Zeno effect one of which is presented at figure 1. and figure 2. . Concretely, at figure 1. it is presented a classical mechanical mechanism (with quasi-statistical or quasi-accidental processes) applied in many real gambling machines. Here a ball B with small dimensions and mass propagates in anti-clockwise direction through a tube with variable cross-sections. In the middle of the left tube part a horizontal two-dimensional square mechanical lattice with $n^2$ equal segments is placed where n represents some natural number. Surface of any lattice segment is much larger than B dimensions. Initially B is placed up in respect to L. For reason of many previous collisions with tube walls (that generate quasi-statistical processes) B has some vertical momentum directed toward L and horizontal momentum of practically any intensity and direction. It implies that B can fall with probability $1/n^2$ at practically any segment of L. Only one segment of L (dark segment at figure 2.) holds

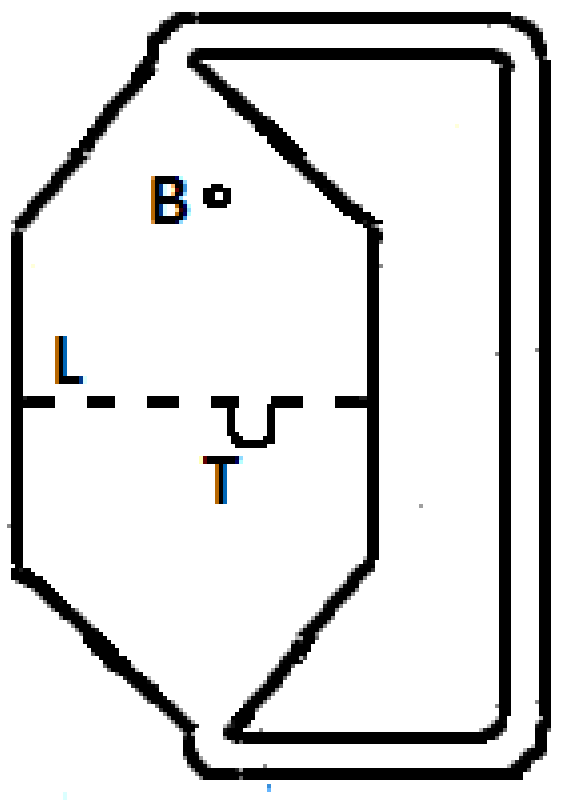

Figure 1.

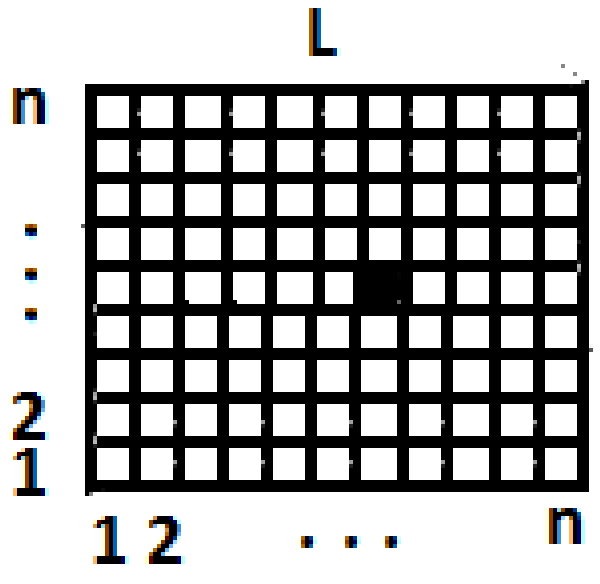

Figure 2.

a trap for B. If B falls in trap it cannot further propagate through tube. In any other situation B goes through L and further propagates through tube. Of course total probability for further propagation of B through tube after one interaction between B and L equals

(2.22) $p(n) = 1 - 1/n^2$ .

Later (by some mechanism) B propagates through down, right and up part of tube and finally new interaction between B and L will be realized. Probability that B n times goes through L then equals

(2.23) $p^n(n) = (1 - 1/n^2)^n$ .

Of course last expression, in limit when n tends to infinity, tends toward 1, i.e.

(2.24) $(1 - 1/n^2)^n \rightarrow 1$ for $n \rightarrow \infty$

which represents a classical Zeno effect (even if here ball eternally moves through tube).

## 3. Quantum effects analogous and similar to quantum Zeno effect

Some authors [3]. [4] considered optical, i.e. polarization detection analogy of quantum Zeno and anti-Zeno effects. We shall now use simplest (Malus law) form of such ideas on the optical, i.e. polarization detection analogy of quantum Zeno and anti-Zeno effects. frequent detections, total probability for propagation of the photon through all analyzers has form of (quasi) exponentially decreasing function.

Consider well-known simple experiment of the single photon polarization. Single non-polarized photon after transition through a polarizer P becomes polarized vertically in initial polarization state |0>. Later this photon quantum dynamically interacts with analyzer A whose polarization axis is rotated for angle θ (that can be any real number, positive, zero or negative) in respect to polarizer vertical polarization axis. Then, according to standard quantum mechanical formalism [6]-[8], probability for propagation of the photon through analyzer with finally changed quantum polarization state $|\theta>$ has the following (Malus law) form

(3.1) $P(|\theta>) = |<\theta|0>|^2 = \cos^2\theta$ .

Exact expression (3.1) can be, for sufficiently small θ, approximated by the following expression

(3.2) $P(|\theta>) = (1 - \theta^2/2)^2 = 1 - \theta^2$ .

More precisely speaking here sufficiently small θ represents any such angle for which function cos θ can be approximated by its second order Taylor expansion approximation. For this reason here all terms that contain θ degree larger than 2 must be neglected according to rules of the approximation calculus. In usual practice, roughly speaking, it can be considered that θ is sufficiently small for

(3.3) $|\theta| < 1/10$

so that

(3.4) $\theta^2 < 1/100$ .

Moreover, if angle θ is sufficiently small then any other angle φ that satisfy condition

(3.5) $|\varphi| < |\theta|$

is sufficiently small too.

Consider now a generalization of the previous experiment presented at figure 3. .
Here also non-polarized single photon after transition through polarizer P becomes initially vertically polarized. Mentioned photon later quantum dynamically interacts with a series of n discretely distant analyzers $A_1$, $A_2$, … , $A_n$ where n is some natural number. Finally there is single photon detector D. If D realizes positive detection it means that photon has been transmitted through any polarizer. Suppose additionally that polarization axis of k-th analyzer (including polarizer as zeroth analyzer) is rotated in respect to polarization axis of previous analyzer for angle $\theta_k$ (that can be any real number, positive, zero or negative) for k= 1, 2, … , n . One concrete choice of all mentioned angles defines one concrete experimental arrangement. In this case final total rotation angle, i.e. angle between polarizer polarization axis and last analyzer polarization axis, equals

(3.6) $\theta(n) = \theta_1 + \theta_2 + \ldots + \theta_n$ .

This angle (3.6) corresponds to the final total polarization state $|\theta(n)>$ .Then, total

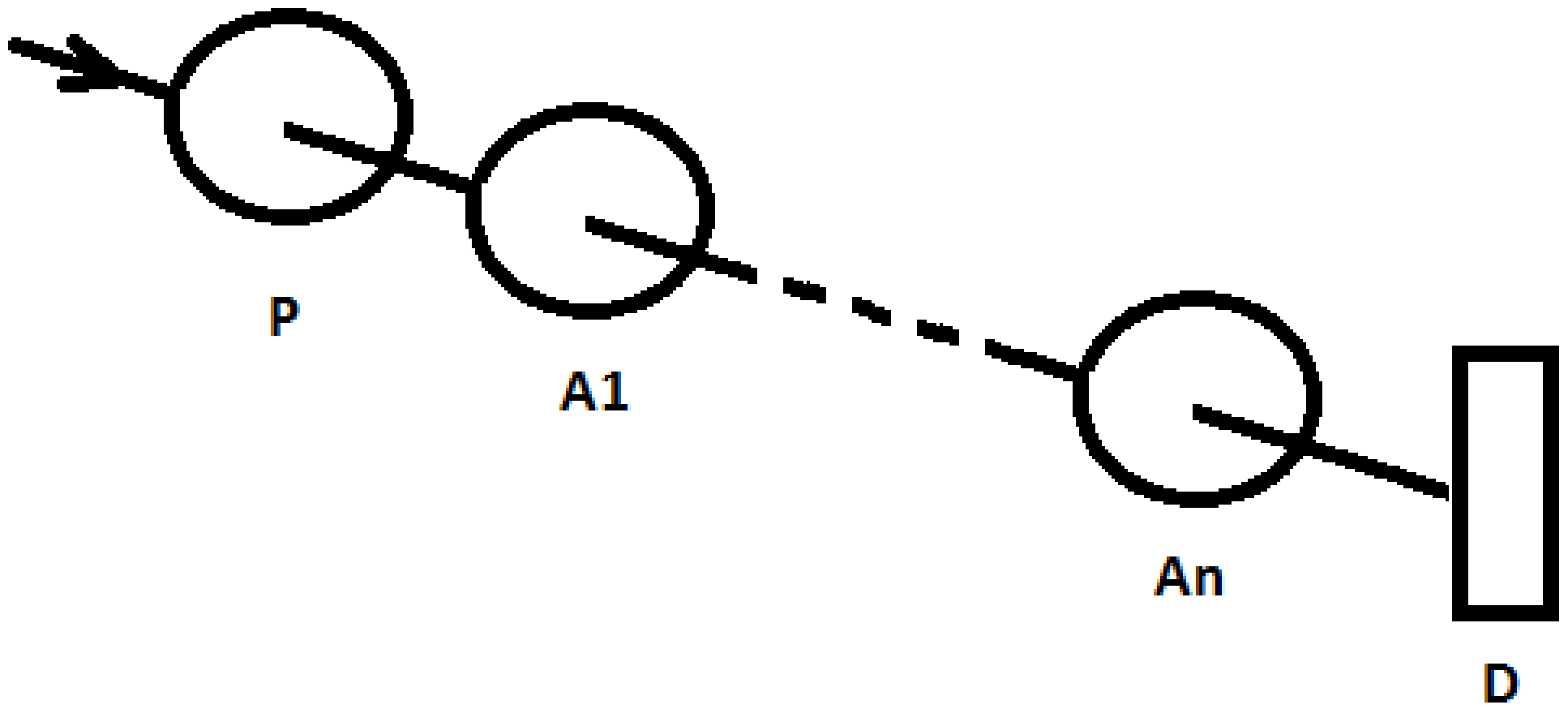

**Figure 3.**

probability for propagation of the photon through all n analyzers equals

(3.7) $P(\theta(n)) = \cos^2\theta_1 \ \cos^2\theta_2 \ \ldots \cos^2\theta_n$ .

It can be added that we shall consider only such non-trivial experimental arrangements in which all members at right side of (3.7) are exactly different from 0 or 1 .

Expressions (3.6) and (3.7) depend of n+1 arguments, number n and n angles $\theta_k$ for k= 1, 2, … ,n . If all values of n angles $\theta_k$ for k= 1, 2, … , n are independent of n it is not hard to see that (in mentioned non-trivial experimental arrangements) expression (3.7) decreases and tends to zero when n increases and tends to infinity. In other situations when all mentioned angles $\theta_k$ for k= 1, 2, … , n are dependent of n corresponding concrete value of (3.7) ( exact or approximated) that depends of n can obtain value different from 0 in limit when n increases and tends to infinity.

Consider, for sufficiently small angles $\theta_k$ for k= 1, 2, … , n, the following approximation

(3.8) $\cos^2\theta_k = (1 - \theta^2_k)$ for k= 1, 2, … ,n

and corresponding approximation of (3.7)

(3.9) $P(\theta(n)) = (1 - \theta^2_1)(1 - \theta^2_2) \ldots (1 - \theta^2_n)$ . .

Consider, for some sufficiently small angle θ (that can be considered as a n independent parameter), the following experimental arrangement

(3.10) $\theta_k = \theta/n$ for k= 1, 2, … ,n . .

(so that angles $\theta_k$ for k= 1, 2, … , n are sufficiently small for any n) when expression (3.6) turns out in

(3.11) $\theta(n) = n(\theta/n) = \theta$ for any n .

Also, according to (3.10), expression (3.9) turns out in

(3.12) $P(\theta) = (1 - (\theta/n)^2)^n = (1 - \theta^2/n^2)^n$ .

Then, in limit when n tends to infinity, (3.12) tends toward 1 , i.e.

(3.13) $P(\theta) \rightarrow 1$ for $n \rightarrow \infty$ .

It definitely means that photon certainly changes initial vertical polarization state |0> in final total polarization state | θ >. It is physically analogous to quantum anti-Zeno effect for non-stable, i.e. decayable quantum systems. (Formally mathematically probability for photon transition through all analyzers has form analogous to probability for total non-decay of the non-stable quantum system by quantum Zeno effect.)

Expression (3.9) , in further approximation for sufficiently small angles $\theta_k$ for k= 1, 2, … , n, can be approximated by the following expression

(3.14) $P(\theta(n)) = (1 - (\theta^2_1 + \theta^2_2 + \ldots + \theta^2_n))$ .

It, according to (3.11), implies

(3.15) $P(\theta(n)) = (1 - n((\theta/n)^2)) = (1 - \theta^2/n)$ .

Then, in limit when n tends to infinity, (3.15) tends toward 1 , i.e.

(3.16) $P(\theta) \rightarrow 1$ for $n \rightarrow \infty$ .

analogous to (3.13).

Consider, for some sufficiently small angle θ (that can be considered as a n independent parameter), next experimental arrangement

(3.17) $\theta_k = (-1)^{k+1} (\theta/n) = (-1)^{k+1} \theta/n$ for k= 1, 2, … ,n

(so that angles $\theta_k$ for k= 1, 2, … , n are sufficiently small for any n) . According to (3.17) expression (3.6) turns out in

(3.18) $\theta(n) = (1 + (-1)^{n+1})/2\ (\theta/n)$

while expression (3.9) turns out in

(3.19) $P(\theta(n)) = (1 - (\theta/n)^2)^n$

numerically identical with (3.12). Then, in limit when n tends to infinity, (3.18) tends to 0, i.e.

(3.20) $\theta(n) \rightarrow 0$ for $n \rightarrow \infty$

and (3.19) tends to 1 , i.e.

(3.21) $P(\theta(n)) \rightarrow 1$ for $n \rightarrow \infty$ .

It definitely means that photon finally certainly does not change initial vertical polarization state |0>. It is physically and formally mathematically analogous to quantum Zeno effect for non-stable, i.e. decayable quantum systems.

Expression (3.19) , in further approximation for sufficiently small angles $\theta_k$ for k= 1, 2, … , n, can be approximated by the following expression

(3.22) $P(\theta(n)) = (1 - (\theta^2_1 + \theta^2_2 + \ldots + \theta^2_n))$ . .

It, according to (3.17), implies

(3.23) $P(\theta(n)) = (1 - n((\theta/n)^2)) = (1 - \theta^2/n)$ . .

Then, in limit when n tends to infinity, (3.23) tends toward 1 , i.e.

(3.24) $P(\theta(n)) \rightarrow 1$ for $n \rightarrow \infty$ .

analogous to (3.21).

Consider, for some sufficiently small angle θ (that can be considered as a n independent parameter), next experimental arrangement

(3.25) $\theta_k = \theta/k$ for k= 1, 2, … ,n

(so that angles $\theta_k$ for k= 1, 2, … , n are sufficiently small for any n) . According to (3.25), in limit $n \rightarrow \infty$, expression (3.6) turns out in well-known divergent harmonic series multiplied with angle θ

(3.26) $\theta(n) = \theta(1 + 1/2 + 1/3 + 1/4 \ldots) = \rightarrow \infty$ for $n \rightarrow \infty$ .

Also, according to (3.25), expression (3.14) turns out in

(3.27) $P(\theta(n) \rightarrow \infty) = 1 - \theta^2 (1 + 1/2^2 + 1/3^2 + 1/4^2 \ldots) = 1 - \theta^2\zeta(2) = 1 - \theta^2\pi^2/6 =$ .
$= 1 - 1.644\theta^2$

where ζ (2) (sum of the hyper harmonic series $(1 + 1/2^2 + 1/3^2 + 1/4^2 \ldots)$) represents Riemann zeta function for argument 2 . It represents very interesting result. Roughly speaking, when detection frequency increases toward infinity then total rotation angle diverges toward infinity and final total polarization state diverge and cannot be determined ("decided"). Nevertheless probability for appearance of any such non-determinable ("non-decidable") final total polarization state is well determined, for chosen sufficiently small θ. Precisely, for appropriately chosen sufficiently small angle θ, angle $1.644^{1/2}\,\theta$ represents sufficiently small angle too, so that expression (3.27) for theoretically predicted probability for propagation of the photon through "all analyzers" is correct. "All analyzers" correspond of course to number n of all analyzers. Suppose "physical limit" where n1 represent finite but very large number of all analyzers in mentioned experimental arrangement (3.25), while n2 represent finite but very large number of all analyzers in mentioned experimental arrangement (3.25) for n2 > n1. Then experimental verification of (3.27) can be realized for n1 analyzers and for n2 analyzers with almost identical positive results. On the other hand θ(n1) and θ(n2) (3.26) in mentioned experiments can be significantly different. Such and similar experimental situations can be called quantum anti-Hamlet effect.

It can be observed that (by formal correspondence $\theta \rightarrow \Delta H\, t/\hbar$ etc.) expressions (3.25)-(3.27) for quantum anti-Hamlet effect for polarization turn out in analogous expressions that characterize quantum anti-Hamlet effect for unstable, decyable quantum system. Of course in this case expression corresponding to (3.27) represents total no-decay probability. In this sense here quantum anti-Hamlet effect for unstable, decyable quantum system is defined.

Consider, for some sufficiently small angle θ (that can be considered as a n independent parameter), next experimental arrangement

(3.28) $\theta_k = (-1)^{k+1}\theta/k^{1/2}$ for k= 1, 2, … ,n .

(so that angles $\theta_k$ for k= 1, 2, … , n are sufficiently small for any n) . According to (3.28), in limit $n \rightarrow \infty$, expression (3.6) turns out in well-known convergent series (equivalent to Dirichlet eta function of argument 1/2 , η(1/2) = 0.604… ) multiplied with angle θ

(3.29) $\theta(n) = \theta\,(1 - 1/2^{1/2} + 1/3^{1/2} - 1/4^{1/2} \ldots.) = \theta\eta(1/2) = 0.604\ldots\ \theta \quad \text{for } n \rightarrow \infty$ .

Also, according to (3.28), expression (3.14) turns out in the following divergent series

(3.30) $P\,(\theta\eta(1/2)) = 1 - \theta^2\,(1 + 1/2 + 1/3 + 1/4 \ldots.) \rightarrow -\infty$

since (1 + 1/2 + 1/3 + 1/4 …. ) represents well-known divergent harmonic series. It represents very interesting result too. Roughly speaking, when detection frequency increases toward infinity then total rotation angle converges and final total polarization state can be determined (“decided”). Nevertheless total probability for appearance of such determinable (“decidable”) final polarization state is divergent or non-determinable (“non-decidable”), for chosen sufficiently small θ. Precisely, for appropriately chosen sufficiently small angle θ, suppose “physical limit” where n1 represent finite but very large number of all analyzers in mentioned experimental arrangement (3.28), while n2 represent finite but very large number of all analyzers in mentioned experimental arrangement (3.28) for n2 > n1. Then experimental verification of (3.29) can be realized for n1 analyzers and for n2 analyzers with almost identical positive results. On the other hand P (θη(1/2)) for n1 and P (θη(1/2)) for n2 (3.30) in mentioned experiments can be significantly different. Moreover, for sufficiently large n1 and n2 expression (3.30) can be negative so that it cannot represent any probability. Generally speaking here expression (3.30) or (3.14) cannot correctly describe probability of propagation of the photon through all analyzers. Such and similar situations can be called quantum Hamlet effect.

It can be observed that mentioned quantum Hamlet effect for polarization cannot be formally transformed corresponding in quantum Hamlet effect for unstable, decyable quantum system by formal correspondence $\theta \rightarrow \Delta H\, t/\hbar$ etc. in expressions (3.28)-(3.30) since in non-relativistic quantum mechanics time t cannot be negative.

## 4. Quantum Zeno and similar effects by photon interference at beam splitter

Consider well-known usual experiment of single photon interference at beam splitter. Here source S emits single photon with well-defined momentum and corresponding wave length that propagates toward beam splitter BS. Immediately after quantum dynamical interaction with BS "reflected" photon in quantum state (trajectory) |R > propagates toward detector DR, while "transmitted" photon in quantum state (trajectory) |T > propagates toward other detector DT so that, immediately before any detection, photon is described by the following quantum superposition state

(4.1) $|ph,a> = i\, a\, |R> + (1 - a^2)^{-1/2} |T>$

where $a$ represents real parameter between 0 and 1 that characterizes quantum dynamical interaction between photon and BS. Then, according to standard quantum mechanical formalism, probability for positive detection of the "reflected" photon by DR equals

(4.2) $P(R) = a^2$

while probability for positive detection of the "transmitted" photon by DT equals

(4.3) $P(T) = 1 - a^2$ .

In case when DR is much closer to BS than DT then negative detection of DR in corresponding time interval implies that photon certainly propagates toward DT.

Mentioned experiment can be simply generalized for a series of the beam splitters BS1, BS2, … , BSn with corresponding characteristic parameters $a_1, a_2, \ldots, a_n$ and corresponding "reflected" photon detectors DR1, DR2, … , DRn where n represents some natural number as it is symbolically presented at figure 4. . Then probability that DRk realizes positive detection equals

(4.4) $P(R,k) = (1 - a^2_1)(1 - a^2_2) \ldots (1 - a^2_{k-1})\, a^2_k$

For k smaller or equal to n, while probability that DT realizes positive detection equals

(4.5) $P(T) = (1 - a^2_1)(1 - a^2_2) \ldots (1 - a^2_n)$

Suppose additionally

(4.6) $a_1 = a_2 = \ldots = a_n = a/n$

when (4.4) and (4.5) turn out in

(4.7) $P(R,k) = (1 - a^2/n^2)^{k-1} (a^2/n^2)$

(for k smaller or equal to n) and

(4.8) $P(T) = (1 - a^2/n^2)^n$ .

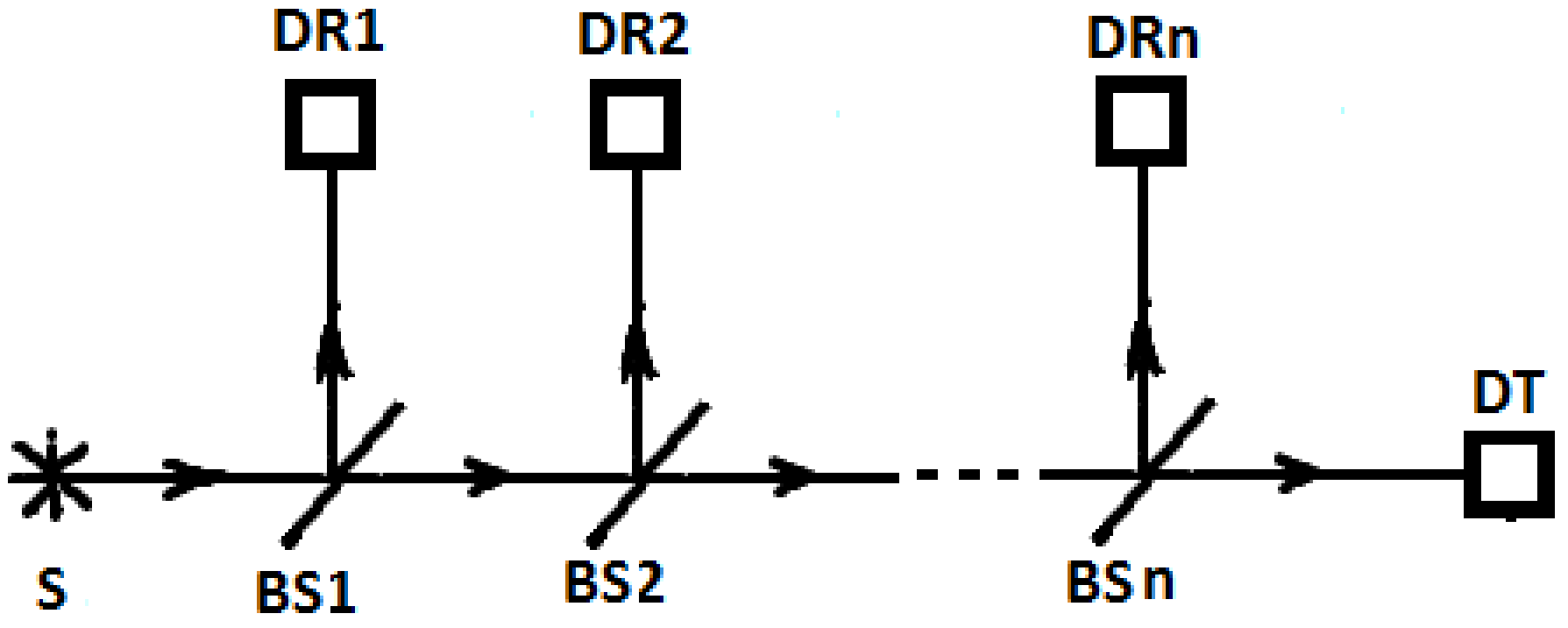

Figure 4.

Then, in limit when n tends to infinity, (4.8) tends toward 1 , i.e.

(4.9) $P(T) \rightarrow 1$ for $n \rightarrow \infty$

which represents quantum Zeno effect. It, strictly speaking, represents clnsequence that, according to (4.6), in limit when n tends to infinity, all beam splitters characteristic parameters tend to zero so that any beam splitter tends to be completely transparent for the photon.

Suppose now

(4.10) $a_1 = a_2 = \ldots = a_n = (1 - a^2/n^2)^{1/2}$

when (4.4) and (4.5) turn out in

(4.11) $P(R,k) = (a^2/n^2)^{k-1} (1 - a^2/n^2)$

(for k smaller or equal to n) and

(4.12) $P(T) = (a^2/n^2)^n$ .

Then, in limit when n tends to infinity, (4.12) tends toward 0 , i.e.

(4.13) $P(T) \rightarrow 0$ for $n \rightarrow \infty$

while

(4.14) $P(R,k) = (a^2/n^2)^{k-1} (1 - a^2/n^2)$

tends to 1 for k=1 and to 0 for k > 1. It represents quantum anti-Zeno effect. Strictly speaking, it represents consequence of the fact that, according to (4.10), in limit when n tends to infinity, all beam splitters characteristic parameters tend to one so that any beam splitter tends to be total mirror for the photon.

Suppose further

(4.15) $a_1 = a_2 = \ldots = a_n = ((1/2)(1 - 2/n^2))^{1/2}$

when (4.4) and (4.5) turn out in

(4.16) $P(R,k) = (1/2 + 1/n^2)^{k-1} ((1/2)(1 - 2/n^2))$

(for k smaller or equal to n) and

(4.17) $P(T) = (1/2 + 1/n^2)^n$ .

Then, in limit when n tends to infinity, (4.17) tends toward 0 , i.e.

(4.19) $P(T) \rightarrow 0$ for $n \rightarrow \infty$

while (4.16) tends toward

(4.20) $P(R,k) = (1/2)^k$

"usual", n-independent exponential law (probability that DRk detects photon exponentially decrease when k increase) characteristic for example by radioactive decays.

## 5. Conclusion

In conclusion the following can be repeated and pointed out. Simply speaking quantum Zeno effect for an unstable quantum system represents total decay probability decrease by frequent decay detections. Analogously simply speaking quantum anti-Zeno effect for an unstable quantum system represents total no-decay probability decrease by frequent decay detections. Also, some authors considered optical, i.e. polarization detection analogy of quantum Zeno and anti-Zeno effects. In this work we consider basic standard quantum mechanical concepts in quantum Zeno and anti-Zeno effect. We use simplest (Malus law) form of mentioned ideas on the photon polarization detection analogy of quantum Zeno and anti-Zeno effects and we suggest some new situations (non-equivalent time durations of the quantum dynamical evolutions between different frequent detections). Correspondingly we define new quantum effects, quantum Hamlet effect (when, by frequent detections, final total polarization angle and state converge but total probability for appearance of this state diverges) and quantum anti-Hamlet effect (when, by frequent detections, final total polarization angle and state diverge but total probability for appearance of this state converges). Also, we consider a rough classical mechanical analogy of Zeno effect. Finally we consider frequent interference of single photon at beam splitters and obtain for different values of reflection and transmission coefficients analogies of quantum Zeno and anti-Zeno effect as well as analogy of usual exponentially decreasing decay law.